\documentclass[aps,prd,article,twocolumn,reprint,superscriptaddress,floatfix]{revtex4-2}
\usepackage{graphicx,amsfonts,xcolor,comment,amsmath,amssymb,float,hyperref,appendix,dcolumn}
\usepackage{bm}
\usepackage{url}
\usepackage{multirow}
\usepackage{soul,ulem}
\preprint{APS/123-QED}

\definecolor{amber}{rgb}{1.0, 0.49, 0.0}

\begin{document}

\title{Search for TeV emission from \textit{spider} millisecond pulsars with HAWC}

\author{R.~Alfaro}
\affiliation{Instituto de F\'{i}sica, Universidad Nacional Autónoma de México, Ciudad de Mexico, Mexico}

\author{E.~Anita-Rangel}
\affiliation{Instituto de Astronom\'{i}a, Universidad Nacional Autónoma de México, Ciudad de Mexico, Mexico}

\author{M.~Araya}
\affiliation{Universidad de Costa Rica, San José 2060, Costa Rica}

\author{J.C.~Arteaga-Velázquez}
\affiliation{Universidad Michoacana de San Nicolás de Hidalgo, Morelia, Mexico}

\author{D.~Avila Rojas}
\affiliation{Instituto de Astronom\'{i}a, Universidad Nacional Autónoma de México, Ciudad de Mexico, Mexico}

\author{H.A.~Ayala Solares}
\affiliation{Temple University, Department of Physics, 1925 N. 12th Street, Philadelphia, PA 19122, USA}

\author{R.~Babu}
\affiliation{Department of Physics and Astronomy, Michigan State University, East Lansing, MI, USA}

\author{P.~Bangale}
\affiliation{Temple University, Department of Physics, 1925 N. 12th Street, Philadelphia, PA 19122, USA}

\author{E.~Belmont-Moreno}
\affiliation{Instituto de F\'{i}sica, Universidad Nacional Autónoma de México, Ciudad de Mexico, Mexico}

\author{A.~Bernal}
\affiliation{Instituto de Astronom\'{i}a, Universidad Nacional Autónoma de México, Ciudad de Mexico, Mexico}

\author{F.~Calore} 
\email{francesca.calore@lapth.cnrs.fr}
\affiliation{Laboratoire d'Annecy-le-Vieux de
Physique Théorique (LAPTh), CNRS, USMB, F-74940 Annecy, France}

\author{T.~Capistrán}
\affiliation{Università degli Studi di Torino, I-10125 Torino, Italy}

\author{A.~Carramiñana}
\affiliation{Instituto Nacional de Astrof\'{i}sica, Óptica y Electrónica, Puebla, Mexico}

\author{S.~Casanova}
\affiliation{Institute of Nuclear Physics Polish Academy of Sciences, PL-31342 IFJ-PAN, Krakow, Poland}

\author{A.L.~Colmenero-Cesar}
\affiliation{Universidad Michoacana de San Nicolás de Hidalgo, Morelia, Mexico}

\author{U.~Cotti}
\affiliation{Universidad Michoacana de San Nicolás de Hidalgo, Morelia, Mexico}

\author{J.~Cotzomi}
\affiliation{Facultad de Ciencias F\'{i}sico Matemáticas, Benemérita Universidad Autónoma de Puebla, Puebla, Mexico}

\author{S.~Coutiño de León}
\affiliation{Dept. of Physics and Wisconsin IceCube Particle Astrophysics Center, University of Wisconsin{\textemdash}Madison, Madison, WI, USA}

\author{E.~De la Fuente}
\affiliation{Departamento de F\'{i}sica, Centro Universitario de Ciencias Exactase Ingenierias, Universidad de Guadalajara, Guadalajara, Mexico}

\author{P.~Desiati}
\affiliation{Dept. of Physics and Wisconsin IceCube Particle Astrophysics Center, University of Wisconsin{\textemdash}Madison, Madison, WI, USA}

\author{N.~Di Lalla}
\affiliation{Department of Physics, Stanford University: Stanford, CA 94305–4060, USA}

\author{R.~Diaz Hernandez}
\affiliation{Instituto Nacional de Astrof\'{i}sica, Óptica y Electrónica, Puebla, Mexico}

\author{M.A.~DuVernois}
\affiliation{Dept. of Physics and Wisconsin IceCube Particle Astrophysics Center, University of Wisconsin{\textemdash}Madison, Madison, WI, USA}

\author{J.C.~Díaz-Vélez}
\affiliation{Dept. of Physics and Wisconsin IceCube Particle Astrophysics Center, University of Wisconsin{\textemdash}Madison, Madison, WI, USA}

\author{K.~Engel}
\affiliation{Department of Physics, University of Maryland, College Park, MD, USA}

\author{T.~Ergin}
\affiliation{Department of Physics and Astronomy, Michigan State University, East Lansing, MI, USA}

\author{C.~Espinoza}
\affiliation{Instituto de F\'{i}sica, Universidad Nacional Autónoma de México, Ciudad de Mexico, Mexico}

\author{K.~Fang} 
\email{kefang@physics.wisc.edu}
\affiliation{Dept. of Physics and Wisconsin IceCube Particle Astrophysics Center, University of Wisconsin{\textemdash}Madison, Madison, WI, USA}

\author{N.~Fraija}
\affiliation{Instituto de Astronom\'{i}a, Universidad Nacional Autónoma de México, Ciudad de Mexico, Mexico}

\author{S.~Fraija}
\affiliation{Instituto de Astronom\'{i}a, Universidad Nacional Autónoma de México, Ciudad de Mexico, Mexico}

\author{J.A.~García-González}
\affiliation{Tecnologico de Monterrey, Escuela de Ingenier\'{i}a y Ciencias, Ave. Eugenio Garza Sada 2501, Monterrey, N.L., Mexico, 64849}

\author{F.~Garfias}
\affiliation{Instituto de Astronom\'{i}a, Universidad Nacional Autónoma de México, Ciudad de Mexico, Mexico}

\author{A.~Galván-Gámez}
\affiliation{Instituto de F\'{i}sica, Universidad Nacional Autónoma de México, Ciudad de Mexico, Mexico }

\author{N.~Ghosh}
\affiliation{Department of Physics, Michigan Technological University, Houghton, MI, USA}

\author{A.~Gonzalez Muñoz}
\affiliation{Instituto de F\'{i}sica, Universidad Nacional Autónoma de México, Ciudad de Mexico, Mexico}

\author{M.M.~González}
\affiliation{Instituto de Astronom\'{i}a, Universidad Nacional Autónoma de México, Ciudad de Mexico, Mexico}

\author{J.A.~González}
\affiliation{Universidad Michoacana de San Nicolás de Hidalgo, Morelia, Mexico}

\author{J.A.~Goodman}
\affiliation{Department of Physics, University of Maryland, College Park, MD, USA}

\author{S.~Groetsch} 
\email{sjgroets@mtu.edu}
\affiliation{Dept. of Physics and Wisconsin IceCube Particle Astrophysics Center, University of Wisconsin{\textemdash}Madison, Madison, WI, USA}

\author{D.~Guevel}
\affiliation{Department of Physics, Michigan Technological University, Houghton, MI, USA}

\author{J.~Gyeong}
\affiliation{Department of Physics, Sungkyunkwan University, Suwon 16419, South Korea}

\author{J.P.~Harding}
\affiliation{Los Alamos National Laboratory, Los Alamos, NM, USA}

\author{S.~Hernández-Cadena}
\affiliation{Tsung-Dao Lee Institute \& School of Physics and Astronomy, Shanghai Jiao Tong University, 800 Dongchuan Rd, Shanghai, SH 200240, China}

\author{I.~Herzog}
\affiliation{Department of Physics and Astronomy, Michigan State University, East Lansing, MI, USA}

\author{D.~Huang}
\affiliation{Department of Physics and Astronomy, University of Delaware, Newark, DE, USA.}

\author{F.~Hueyotl-Zahuantitla}
\affiliation{Universidad Autónoma de Chiapas, Tuxtla Gutiérrez, Chiapas, México}

\author{P.~Hüntemeyer}
\affiliation{Department of Physics, Michigan Technological University, Houghton, MI, USA}

\author{A.~Iriarte}
\affiliation{Instituto de Astronom\'{i}a, Universidad Nacional Autónoma de México, Ciudad de Mexico, Mexico}

\author{S.~Kaufmann}
\affiliation{Universidad Politecnica de Pachuca, Pachuca, Hgo, Mexico}

\author{D.~Kieda}
\affiliation{Department of Physics and Astronomy, University of Utah, Salt Lake City, UT, USA}

\author{K.~Leavitt}
\affiliation{Department of Physics, Michigan Technological University, Houghton, MI, USA}

\author{W.H.~Lee}
\affiliation{Instituto de Astronom\'{i}a, Universidad Nacional Autónoma de México, Ciudad de Mexico, Mexico}

\author{H.~León Vargas}
\affiliation{Instituto de F\'{i}sica, Universidad Nacional Autónoma de México, Ciudad de Mexico, Mexico}

\author{A.L.~Longinotti}
\affiliation{Instituto de Astronom\'{i}a, Universidad Nacional Autónoma de México, Ciudad de Mexico, Mexico}

\author{G.~Luis-Raya}
\affiliation{Universidad Politecnica de Pachuca, Pachuca, Hgo, Mexico}

\author{K.~Malone}
\affiliation{Los Alamos National Laboratory, Los Alamos, NM, USA}

\author{S.~Manconi} 
\email{manconi@lpthe.jussieu.fr}
\affiliation{Laboratoire d'Annecy-le-Vieux de Physique Théorique (LAPTh), CNRS, USMB, F-74940 Annecy, France}
\affiliation{Sorbonne Universit\'e \& Laboratoire de Physique Th\'eorique et Hautes \'Energies (LPTHE),CNRS, 4 Place Jussieu, Paris, France}

\author{O.~Martinez}
\affiliation{Facultad de Ciencias F\'{i}sico Matemáticas, Benemérita Universidad Autónoma de Puebla, Puebla, Mexico}

\author{J.~Martínez-Castro}
\affiliation{Centro de Investigaci\'on en Computaci\'on, Instituto Polit\'ecnico Nacional, M\'exico City, M\'exico.}

\author{J.A.~Matthews}
\affiliation{Dept of Physics and Astronomy, University of New Mexico, Albuquerque, NM, USA}

\author{P.~Miranda-Romagnoli}
\affiliation{Universidad Autónoma del Estado de Hidalgo, Pachuca, Mexico}

\author{J.A.~Morales-Soto}
\affiliation{Universidad Michoacana de San Nicolás de Hidalgo, Morelia, Mexico}

\author{M.~Mostafá}
\affiliation{Temple University, Department of Physics, 1925 N. 12th Street, Philadelphia, PA 19122, USA}

\author{M.~Najafi}
\affiliation{Department of Physics, Michigan Technological University, Houghton, MI, USA}

\author{L.~Nellen}
\affiliation{Instituto de Ciencias Nucleares, Universidad Nacional Autónoma de Mexico, Ciudad de Mexico, Mexico}

\author{R.~Noriega-Papaqui}
\affiliation{Universidad Autónoma del Estado de Hidalgo, Pachuca, Mexico}

\author{N.~Omodei}
\affiliation{Department of Physics, Stanford University: Stanford, CA 94305–4060, USA}

\author{M.~Osorio-Archila}
\affiliation{Instituto de Astronom\'{i}a, Universidad Nacional Autónoma de México, Ciudad de Mexico, Mexico}

\author{E.~Ponce}
\affiliation{Facultad de Ciencias F\'{i}sico Matemáticas, Benemérita Universidad Autónoma de Puebla, Puebla, Mexico}

\author{Y.~Pérez Araujo}
\affiliation{Instituto de F\'{i}sica, Universidad Nacional Autónoma de México, Ciudad de Mexico, Mexico}

\author{C.D.~Rho}
\affiliation{Department of Physics, Sungkyunkwan University, Suwon 16419, South Korea}

\author{A.~Rodriguez Parra}
\affiliation{Universidad Michoacana de San Nicolás de Hidalgo, Morelia, Mexico}

\author{D.~Rosa-González}
\affiliation{Instituto Nacional de Astrof\'{i}sica, Óptica y Electrónica, Puebla, Mexico}

\author{M.~Roth}
\affiliation{Los Alamos \ Laboratory, Los Alamos, NM, USA}

\author{H.~Salazar}
\affiliation{Facultad de Ciencias F\'{i}sico Matemáticas, Benemérita Universidad Autónoma de Puebla, Puebla, Mexico}

\author{A.~Sandoval}
\affiliation{Instituto de F\'{i}sica, Universidad Nacional Autónoma de México, Ciudad de Mexico, Mexico}

\author{M.~Schneider}
\affiliation{Department of Physics, University of Maryland, College Park, MD, USA}

\author{J.~Serna-Franco}
\affiliation{Instituto de F\'{i}sica, Universidad Nacional Autónoma de México, Ciudad de Mexico, Mexico}

\author{M.~Shin}
\affiliation{Department of Physics, Sungkyunkwan University, Suwon 16419, South Korea}

\author{Y.~Son}
\affiliation{University of Seoul, Seoul, Rep. of Korea}

\author{R.W.~Springer}
\affiliation{Department of Physics and Astronomy, University of Utah, Salt Lake City, UT, USA}

\author{O.~Tibolla}
\affiliation{Universidad Politecnica de Pachuca, Pachuca, Hgo, Mexico}

\author{K.~Tollefson}
\affiliation{Department of Physics and Astronomy, Michigan State University, East Lansing, MI, USA}

\author{I.~Torres}
\affiliation{Instituto Nacional de Astrof\'{i}sica, Óptica y Electrónica, Puebla, Mexico}

\author{F.~Ureña-Mena}
\affiliation{Instituto Nacional de Astrof\'{i}sica, Óptica y Electrónica, Puebla, Mexico}

\author{E.~Varela}
\affiliation{Facultad de Ciencias F\'{i}sico Matemáticas, Benemérita Universidad Autónoma de Puebla, Puebla, Mexico}

\author{X.~Wang}
\affiliation{Department of Physics, Missouri University of Science and Technology, Rolla, MO, US}

\author{Z.~Wang}
\affiliation{Department of Physics, Missouri University of Science and Technology, Rolla, MO, US}

\author{H.~Wu} 
\email{hwu298@wisc.edu}
\affiliation{Dept. of Physics and Wisconsin IceCube Particle Astrophysics Center, University of Wisconsin{\textemdash}Madison, Madison, WI, USA}

\author{S.~Yu}
\affiliation{Department of Physics, Pennsylvania State University, University Park, PA, USA}

\author{X.~Zhang}
\affiliation{Institute of Nuclear Physics Polish Academy of Sciences, PL-31342 IFJ-PAN, Krakow, Poland}

\author{H.~Zhou}
\affiliation{Tsung-Dao Lee Institute \& School of Physics and Astronomy, Shanghai Jiao Tong University, 800 Dongchuan Rd, Shanghai, SH 200240, China}

\author{C.~de León}
\affiliation{Universidad Michoacana de San Nicolás de Hidalgo, Morelia, Mexico}

\begin{abstract}
Millisecond pulsars (MSPs) are observed to emit multi-wavelength radiation, from radio to GeV. Spider MSPs, which interact with their low-mass companion in close orbit (orbital periods $< 1$~day), may lead to strong intrabinary shocks that can further accelerate electron and positron pairs produced in the magnetosphere, possibly emitting very-high-energy (0.1--100~TeV; VHE) photons through inverse Compton scattering. Using 2565 days of HAWC Pass~5 data, we search for VHE emission from spider MSPs and  present upper limits on individual sources. We also perform a stacking analysis to examine whether the two sets of spider systems, classified as redbacks and black widows depending on the companion mass, exhibit different spectral properties. Our study places constraints on TeV emission from MSPs and suggests that they are unlikely to contribute significantly to the Galactic diffuse emission at TeV and higher energies. 

\end{abstract}

\maketitle 

\section{Introduction}
Pulsars are fast-rotating, magnetized neutron stars that are known to emit from radio to gamma rays.
323 pulsars \footnote{https://confluence.slac.stanford.edu/x/5Jl6Bg} are currently identified at GeV gamma-ray energies by the Large Area Telescope (LAT) aboard the \textit{Fermi} satellite \cite{Fermi-LAT:2023zzt}. Among them, a few young pulsars have been also observed to emit up to TeV energies \cite{MAGIC:2015ggt,HESS:2023sxo}. 
The gamma-ray observations have prompted investigations into several key aspects of particle acceleration in these systems, including the mechanisms that produce pulsed and phase-averaged emission, the accelerated particle species, and their possible contribution to Galactic cosmic rays and diffuse emissions  \cite{Lopez-Coto:2022igd,Liu:2022hqf,Fang:2022fof,Amato:2024dss}. In particular, the termination shock of the pulsar wind observed in young (${\sim}10$ kyr) and middle-aged (${\sim}100$ kyr) pulsars is an ideal environment to accelerate the electrons and positrons produced in the pulsar magnetosphere. These particles could then form TeV halos around the pulsar \cite{HAWC:2017kbo,HAWC:2024scl,HESS:2023sbf, Albert:2025gwm} and propagate through the Galaxy, contributing to the local fluxes of cosmic rays observed by AMS-02 (see e.g. \cite{Evoli:2020szd,Manconi:2020ipm,Orusa:2024ewq}). 

Apart from young and middle-aged pulsars, more than half of the population of GeV-emitting pulsars consists of older neutron stars, the so-called millisecond pulsars (MSPs) which spin with periods in the millisecond range and are most often found in binary systems. 
Millisecond pulsars are also observed to emit multi-wavelength radiation from radio to GeV energies. However, no firm detection of MSPs at TeV energies has been established yet~\cite{2022ASSL..465...57H} despite deep searches being carried out on promising targets (e.g. PSR J0218+4232 \cite{acciari_search_2021}). 
Some hints of high-energy gamma rays from globular clusters, systems believed to be dominated by MSPs, have been previously reported \cite{Song:2021zrs,2025A&A...696L..11S}.
A study with HAWC data \cite{hawc_collaboration_absence_2025} found that pulsar halos at TeV energies are not a common feature of MSPs, at least when considering similar properties (e.g., ~efficiency of conversion of the spin-down power to gamma rays) with respect to the observed halos of TeV emission around middle-aged pulsars. This study could not confirm previous claims of TeV halo emission from MSPs \cite{Hooper:2021kyp}.
Nevertheless, pulsar halos are not the only phenomenon that could produce gamma-ray emission at TeV energies in MSPs. 

A growing class of gamma-ray pulsars, involving non-accreting MSPs in compact binary systems, is emerging in GeV catalogs \cite{Fermi-LAT:2023zzt}. These systems, with orbital periods shorter than a day, are classified as \textit{black widows} or \textit{redbacks} \cite{roberts_new_2011} depending on the mass of the companion; collectively, they are often referred to as \textit{spiders} \cite{Koljonen:2025sxl}. The MSP winds in such binaries are thought to interact with the outflows from the low-mass companions, leading to strong intrabinary shocks that can further accelerate electron and positron pairs produced in the magnetosphere, possibly producing VHE emission through inverse Compton scattering in the photon field supplied by the low-mass companion \cite{Wadiasingh:2021wcj,Cortes:2022tzh,Sim:2024kxi,Richard-Romei:2024nje}.

Redbacks have heavier companions, with masses of 0.1--0.4~$M_{\odot}$, and black widows have companions with masses less than 0.1~$M_{\odot}$. Redbacks are expected to be more efficient emitters at TeV energies, given the higher companion mass, which could provide denser targets for producing inverse Compton scattering gamma-ray radiation \cite{Wadiasingh:2021wcj}. Also, there are hints that redbacks are more efficient emitters at X-ray energies with respect to black widows, although it is not clear whether the X-ray and gamma-ray emission originates from the same mechanism \cite{Hui:2019pin}.

Based on the mechanism outlined above, certain theoretical models \cite{Wadiasingh:2021wcj,Richard-Romei:2024nje} suggest the possibility of TeV emission. However, large theoretical uncertainties (e.g., in shock geometry and acceleration efficiency) are at play, preventing precise predictions for the intensity and spectral properties of this emission. 
Multi-TeV observations of spider systems thus have the potential to unveil, for the first time, TeV gamma-ray emission from MSPs, informing us about the acceleration mechanism at intrabinary shocks and the properties of the particle population accelerated \cite{Wadiasingh:2021wcj}.

Using 2,565 days of data from the High-Altitude Water Cherenkov (HAWC) Gamma-Ray Observatory, in this paper we provide the first search for VHE emission from a sample of spider MSPs through several approaches, inspired by the work carried out in \cite{Albert:2025gwm,hawc_collaboration_absence_2025} and with a number of improvements. Specifically, we examine whether the two sets of spider systems, the redbacks and black widows, exhibit different emission properties. Given the expected faint emission from each individual source, a joint likelihood technique is used to search for a cumulative signal from the two different sets. A dedicated region of interest (ROI) analysis using multi-source fitting is also carried out for a few targets that present mild excesses of emission in the standard analysis used in the first step. 
In the absence of a significant signal from the stacked analysis of HAWC data from all MSPs in our sample, as well as from the dedicated single-source analysis, we derive upper limits on the average flux in the energy range 0.32--100~TeV. 

This paper is organized as follows: In Sec.~\ref{sec:method} we present the data selection, MSP sample definition, and methodology adopted to perform both the stacked and individual-source analysis.  Results are compiled in Sec.~\ref{sec:results} first for the stacking analysis, and then for the individual-source fits. We present the analysis of systematic uncertainties in Sec.~\ref{sec:syst}.
Finally, we discuss the implications of our results and conclude in Sec.~\ref{sec:conclusions}.

\section{Method}\label{sec:method}
\subsection{Telescope and data selection}
The HAWC Observatory is located within the Pico de Orizaba National Park  in the state of Puebla, Mexico, at a latitude of $19^\circ$ North and an elevation of 4,100 meters. It consists of 300 water Cherenkov detectors and features an instantaneous field of view that covers approximately 15\% of the celestial sphere, allowing it to survey nearly two-thirds of the sky every 24 hours \cite{abeysekara2023}. The HAWC reconstruction and analysis algorithms are optimized for primary gamma rays in the energy range of approximately 300~GeV to several hundred TeV. To improve its sensitivity at the highest energies, HAWC has been upgraded with a sparse outrigger array consisting of 350 smaller water Cherenkov detectors \cite{joshiHAWCHighEnergy2017}. Data from the outrigger array are not used in this analysis.

This study uses the 2,565 days of data from HAWC's \texttt{Pass 5.final (5.f)}, which is the latest version of the algorithms used to reconstruct data from the primary detector of the HAWC instrument \cite{albert_performance_2024}. We also used the standard pass 5 gamma/hadron separation cuts. Our analysis uses the Multi-Mission Maximum Likelihood ({\it 3ML}) software framework \citep{vianello_multi-mission_2017} with the HAWC Accelerated Likelihood (HAL) plugin \citep{HAL_plugin}, using ROOT\cite{ROOT_NIMA_1997} Minuit2 as the main minimizer. 

\begin{table}[H]
\centering
\begin{tabular}{ccccccc}\hline
Energy Bin & Energy Range & Pivot Energy \\
 & [TeV] & [TeV]\\ \hline
full & 0.32--100.00 & 5.62\\
1 & 0.32--1.00 & 0.56\\
2 & 1.00--3.16 & 1.78\\
3 & 3.16--10.00 & 5.62\\
4 & 10.00--31.62 & 17.78\\
5 & 31.62--100.00 & 56.23 \\ \hline
\end{tabular}
\caption{Energy binning adopted in this study, including energy intervals and medians (i.e.,~pivot energies).}
\label{tab:paras_e}
\end{table}

We employ five quasi-differential energy bins from 0.32 to 100 TeV, with their median energies and boundaries listed in Table \ref{tab:paras_e}. The pivot energy, $E_{\rm piv}$, used in the stacking analysis is set to be the median of the energy bin in a logarithmic scale. This choice is the same as in the previous study on MSP TeV Halos \cite{hawc_collaboration_absence_2025}.

\subsection{Millisecond pulsars sample}
The list of target spider MSPs was obtained from the University of Maryland (UMD) list of known Galactic MSPs \footnote{\url{https://www.astro.umd.edu/~eferrara/pulsars/GalacticMSPs.txt}, version of 03/05/2024}, which contains a continuously updated list of Galactic MSPs detected at different wavelengths.  
We first filtered this list and selected for sources where the orbital period of the companion is less than one day ($P_b<1$). We obtained 78 sources. Requiring the sources to be in HAWC's field of view ($-26^{\circ} < \delta < 64^{\circ}$), we reduce this set to 53 sources. To complement our selection, we also searched for spider MSPs in the Australia Telescope National Facility (ATNF) catalog (version 2.1.1, 03/04/2024) \cite{ATNF2004}, finding one additional pulsar (J1946+2052). This first selection, therefore, includes 54 spider MSPs.

To study the association between a possible detection of TeV gamma rays and the spider MSP properties, we further separate the sources into redbacks and black widows. 
To this end, we need an estimate of the companion mass. This is provided in the ATNF catalog for 49 sources among the initial set of 54. The companion mass of one additional source (J1705-1903) is provided in \cite{morello_high_2019}. Using a mass threshold of $0.1 M_{\odot}$, 18 redbacks ($M>0.1 M_{\odot}$) and 32 black widows ($M<0.1 M_{\odot}$) are selected. In the categorization process, sources lacking a determined companion mass were excluded.

Finally, to minimize the influence of nearby bright extended sources and crowded regions, only sources that are $\geq 2 ^\circ$ away from any TeVCat \footnote{https://tevcat.org/} sources are included in the analysis, further reducing our spider sample to 43 MSPs: 15 redbacks and 28 black widows. The $2 ^\circ$ deg cut was determined by testing a range of radii to find a balance between two competing requirements: 1. minimizing TS excess caused by known bright TeV emitters; 2. maintaining a sufficiently large sample size to ensure the statistical validity of the study.

\begin{figure*}[t]
\centering
\includegraphics[width = 0.9\textwidth]{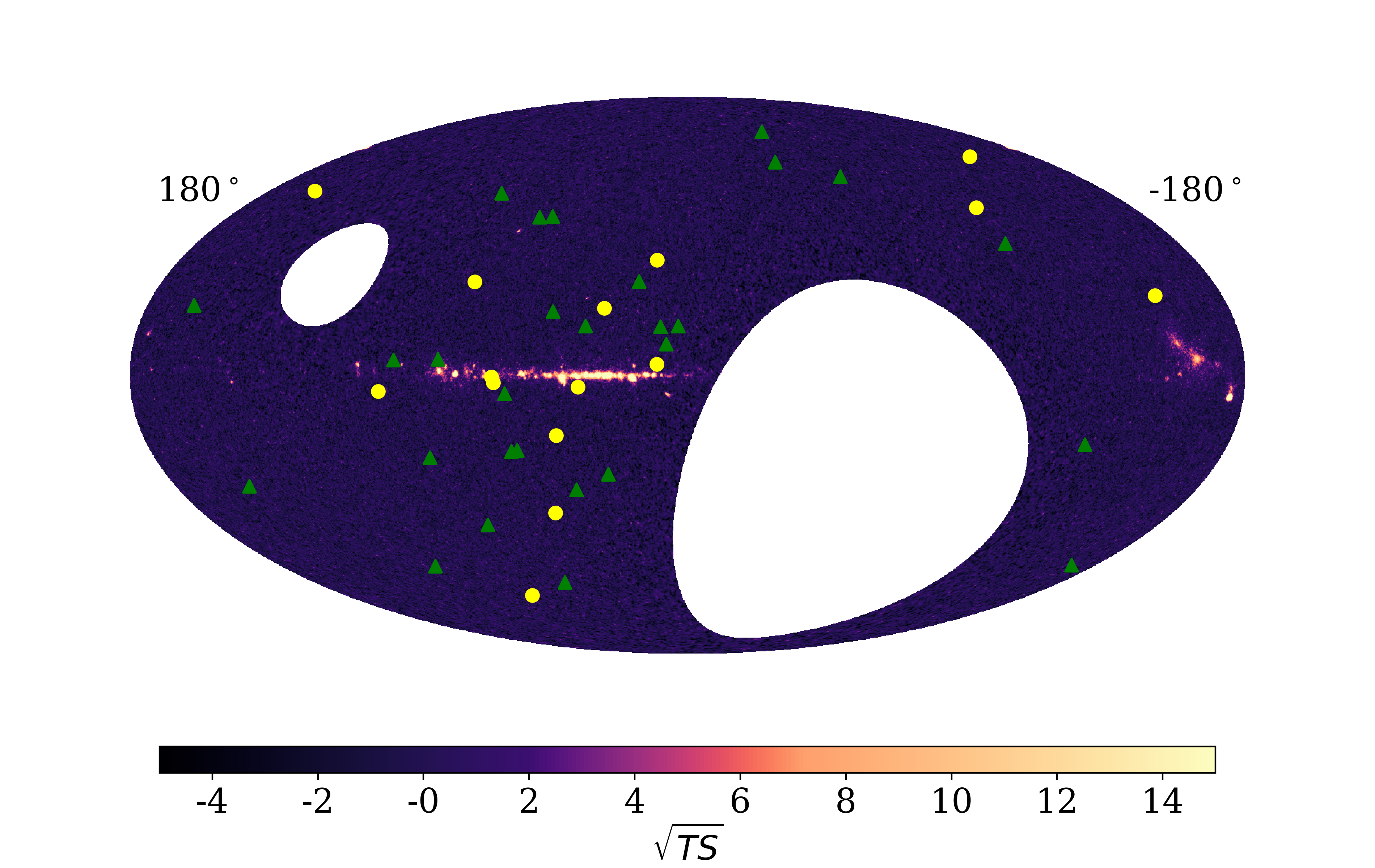}
\caption{Locations of the 43 spider MSPs in our sample in Galactic coordinates. The 15 redbacks are shown as yellow circles and the 28 black widows are shown as green triangles. The color map indicates the significance of point-like sources in HAWC's field-of-view. The significance is approximated as the square root of the test statistics defined in equation~\eqref{eqn:TS} assuming a spectral index $p=2.7$ \cite{wilks1938}.}
\label{fig:skymap}
\end{figure*}

The complete sample of spider MSPs for which we search for TeV emission in HAWC data thus consists of 43 MSPs. Figure \ref{fig:skymap} shows the spatial distribution of these MSPs on the skymap. The spider MSP subsets are highlighted with different markers. The source names, groups and locations are reported in the first four columns of Table~\ref{tab:paras_psr}.

\begin{table*}
\centering
\begin{tabular}{cccccccc}
\hline
\multicolumn{1}{c}{NAME} & \multicolumn{1}{c}{Group} & \multicolumn{1}{c}{RAJD} & \multicolumn{1}{c}{DECJD} & \multicolumn{1}{c}{TS} & \multicolumn{1}{c}{90\% UL}\\
\multicolumn{1}{c}{} & \multicolumn{1}{c}{} & \multicolumn{1}{c}{[$^\circ$]} & \multicolumn{1}{c}{[$^\circ$]} & \multicolumn{1}{c}{} & \multicolumn{1}{c}{[$1 / \rm (cm^2\cdot TeV\cdot s)$]}\\
\hline
J0023+0923 & black widow & 5.82 & 9.44 & 0.69 & $2.36 \times 10^{-14}$ \\
J0251+2606 & black widow & 42.83 & 26.17 & 1.96 & $1.04 \times 10^{-15}$ \\
J0312-0921 & black widow & 48.00 & -9.37 & 0.00 & $1.82 \times 10^{-14}$ \\
J0610-2100 & black widow & 92.52 & -20.97 & 0.00 & $9.30 \times 10^{-14}$ \\
J0636+5128 & black widow & 99.07 & 51.00 & 0.00 & $2.83 \times 10^{-14}$ \\
J0751+1807 & redback & 117.79 & 18.16 & 0.00 & $9.85 \times 10^{-15}$ \\
J0952-0607 & black widow & 148.08 & -6.14 & 0.00 & $1.88 \times 10^{-14}$ \\
J1012+5307 & redback & 153.22 & 53.13 & 0.00 & $2.16 \times 10^{-14}$ \\
J1023+0038 & redback & 155.97 & 0.65 & 0.56 & $2.61 \times 10^{-14}$ \\
J1048+2339 & redback & 162.14 & 23.64 & 0.00 & $1.14 \times 10^{-14}$ \\
J1221-0633 & black widow & 185.36 & -6.60 & 0.00 & $1.38 \times 10^{-14}$ \\
J1301+0833 & black widow & 195.40 & 8.58 & 0.00 & $9.08 \times 10^{-15}$ \\
J1317-0157 & black widow & 199.41 & -1.99 & 0.00 & $1.23 \times 10^{-14}$ \\
J1544+4937 & black widow & 235.96 & 49.64 & 0.00 & $2.37 \times 10^{-14}$ \\
J1622-0315 & redback & 245.77 & -3.18 & 0.89 & $3.50 \times 10^{-14}$ \\
J1627+3219 & black widow & 246.99 & 32.33 & 0.57 & $2.07 \times 10^{-14}$ \\
J1630+3550 & black widow & 247.75 & 35.93 & 0.00 & $1.55 \times 10^{-14}$ \\
J1653-0158 & black widow & 253.43 & -2.01 & 0.28 & $2.49 \times 10^{-14}$ \\
J1705-1903 &  & 256.37 & -19.20 & 0.35 & $1.01 \times 10^{-13}$ \\
J1719-1438 & black widow & 259.85 & -14.62 & 0.00 & $3.71 \times 10^{-14}$ \\
J1731-1847 & black widow & 262.79 & -18.76 & 0.00 & $4.10 \times 10^{-14}$ \\
J1738+0333 & redback & 264.75 & 3.52 & 0.00 & $1.44 \times 10^{-14}$ \\
J1757-1854 & redback & 269.21 & -18.95 & 1.02 & $1.14 \times 10^{-13}$ \\
J1805+0615 & black widow & 271.45 & 6.30 & 0.08 & $1.88 \times 10^{-14}$ \\
J1810+1744 & black widow & 272.64 & 17.70 & 0.31 & $1.72 \times 10^{-14}$ \\
J1816+4510 & redback & 274.05 & 45.17 & 0.00 & $1.92 \times 10^{-14}$ \\
J1906+0055 & redback & 286.65 & 0.83 & 0.93 & $2.70 \times 10^{-14}$ \\
J1952+2630 & redback & 298.15 & 26.45 & 0.00 & $6.86 \times 10^{-14}$ \\
J1957+2516 & redback & 299.29 & 25.25 & $2.19 \times 10^{-4}$ & $7.64 \times 10^{-14}$ \\
B1957+20 & black widow & 299.91 & 20.81 & 0.00 & $5.67 \times 10^{-14}$ \\
J2006+0148 & redback & 301.66 & 1.80 & 0.00 & $1.71 \times 10^{-14}$ \\
J2017-1614 & black widow & 304.42 & -16.24 & 0.36 & $6.89 \times 10^{-14}$ \\
J2047+1053 & black widow & 311.79 & 10.84 & 0.00 & $1.12 \times 10^{-14}$ \\
J2051-0827 & black widow & 312.77 & -8.45 & 2.42 & $5.31 \times 10^{-14}$ \\
J2052+1219 & black widow & 313.19 & 12.29 & 0.00 & $1.06 \times 10^{-14}$ \\
J2055+3829 & black widow & 304.72 & 43.65 & 0.00 & $2.15 \times 10^{-14}$ \\
J2115+5448 & black widow & 318.76 & 54.78 & 0.00 & $3.78 \times 10^{-14}$ \\
J2129-0429 & redback & 322.40 & -4.47 & 0.09 & $1.39 \times 10^{-15}$ \\
J2214+3000 & black widow & 333.72 & 30.01 & 3.28 & $2.67 \times 10^{-14}$ \\
J2215+5135 & redback & 333.97 & 51.58 & 0.00 & $3.15 \times 10^{-14}$ \\
J2234+0944 & black widow & 338.68 & 9.78 & 1.92 & $2.44 \times 10^{-14}$ \\
J2256-1024 & black widow & 344.00 & -10.40 & 0.07 & $4.15 \times 10^{-14}$ \\
J2339-0533 & redback & 354.91 & -5.58 & 0.19 & $2.99 \times 10^{-14}$ \\
\hline
\end{tabular}
\caption{A list of all MSPs in the analysis with their individual best-fit normalization and full energy range 90\% upper limits. Results of J1952+2630, J1957+2516 and B1957+20 are obtained by multi-source fitting along with nearby sources, replacing their minimal source fit results. dK is the average absolute uncertainty of the best-fit normalization.}
\label{tab:paras_psr}
\end{table*}

\medskip

\subsection{Minimal source fit}
We first fit each source in the sample and evaluate its individual detection significance. 

The  gamma-ray flux $\Phi$ of a source is described by a powerlaw spectral model:
\begin{equation}
    \Phi=K \left(\frac{E}{E_{\mathrm{piv}}}\right)^{-p},
\end{equation}
where the spectral index $p$ is set to the mean index of the 2HWC point sources, $p = 2.7$ \cite{2HWC2017}, $K$ is the normalization factor in units of $1 / \rm (cm^2\ TeV\ s)$, and the pivot energy $E_{\mathrm{piv}}$ is defined in Table~\ref{tab:paras_e}. In all fits, $K$ is constrained to be non-negative.

The best-fit parameters are found by maximizing the Poisson likelihood:
\begin{equation}
    {\cal L}(K) = \sum_{j=1}^{N} \log \left\{ \frac{[B_j + S_j(K,j)]^{D_j}}{D_j!} \right\} - [B_j + S_j(K,j)],
\end{equation}
where $B_j$ is the number of background events in the pixel $j$ determined by the HAWC background estimation algorithm \cite{abeysekara_observation_2017}, $S_j$ is the number of events under the assumed model, comprised only by a point-source at the position of each MSP, and $D_j$ is the observed counts. 

The Test Statistic (TS) is defined as twice the logarithm of the likelihood ratio when fitting the data with and without a source model:
\begin{equation}\label{eqn:TS}
    {\rm TS} \equiv2 \ln ({\cal L}(K) / {\cal L} (K = 0))\,.
\end{equation}
The results of the individual fits are the source TS, the best-fit normalization, and uncertainties. In the absence of a significant detection, the 90\% credible upper limits (90\% UL) for each source in the sample is computed by performing a Bayesian Posterior Sampling with a uniform prior on {\it K} and taking the 90\% linear quantile of the sample inverse cumulative distribution function.

We stress that the powerlaw spectral model is a simple model we use as a first guess for the possible TeV emission from MSPs. As reported below, we do not find any significant detection of such emission within HAWC data, and we test the systematic uncertainties connected to the spectral index choice in Section~\ref{sec:syst}. In the case of a future detection, specific spectral shapes suggested by MSP spider models such as log-parabola shapes \cite{Wadiasingh:2021wcj} could be tested directly on the data with similar approaches to what is discussed here.

\subsection{Multi-source fit}\label{method:multi}
For each individual source that presents a TS $>4$ in the minimal source fit, we then perform a multi-source maximum likelihood fit analysis in the respective ROI. The multi-source fit approach accounts for contaminating emission from nearby identified gamma-ray sources and diffuse gamma-ray emission in regions of low Galactic latitude. The previously identified gamma-ray sources used in this multi-source fit are from the upcoming fourth HAWC catalog of gamma-ray sources (4HWC) \cite{groetsch_preliminary_2023}. These sources were identified using an automated source-finding pipeline based on the multi-source likelihood fit and a likelihood ratio test using Wilks' theorem and iterative nested models. 

First, a number of seed point sources are iteratively added to the model. The initial location of each of these sources is determined by the most significant pixel in the residual map generated by subtracting the existing model from the HAWC data map. For the first point source, this map is the same as the HAWC data map shown in Figure \ref{fig:skymap}. These sources are added until the TS computed from the likelihood ratio is below 25. Then, in order of significance, each seed is tested for angular extension by substituting a point-like spatial model with a symmetric Gaussian distribution. If the model shows a TS of 16 or greater, the extended assumption is accepted. Any source that falls below a TS of 25 is pruned from the model, and the overall model is refit to find the new maximum likelihood parameters. This process is repeated until all seeds have been tested or pruned. All surviving seed point sources are then tested for spectral curvature by using a log-parabola spectral model instead of a powerlaw. The same thresholds for acceptance and pruning are used. Then, all surviving models are accepted and refit to find the final maximum likelihood modeling of the data.

This list of sources represents the most recent and sensitive all-sky survey of the gamma-ray sky in HAWC data.

To model the MSPs, a point-like source with a simple powerlaw spectral model is added at the location of the MSP. 
The location of the MSP is fixed. Extended sources are added to the locations of nearby 4HWC sources, and their locations are allowed to vary by up to one degree in right ascension and declination. The spectral parameters, including flux normalization, index, and curvature of the MSPs and 4HWC sources are left free where applicable. The fit is performed using a smaller ROI centered on the MSP with a circular radius of four degrees, which is four times HAWC’s coarsest angular resolution. The radius is chosen to fully include the expected emission while minimizing contamination from nearby sources. The reported TS of the MSP is computed by the likelihood ratio obtained from fits with and without the model corresponding to the location of the MSP.

\subsection{Stacking analysis}
In addition to searching for emission from individual MSPs, we used a stacking technique to assess the statistical evidence of the cumulative emission of different source groups. 
Specifically, we sum the individual-source likelihoods, $\cal L$$_{i}$, to find the stacked TS: 
\begin{equation}
    \ln{\cal L} (K) = \sum_i \ln {\cal L}_i (K)\,,
\end{equation}
where $K$ is the best-fit normalization of the stacked sample assuming the same powerlaw model for all sources. In other words, all sources are assigned equal weight in this analysis. 

Alternative weighting schemes, for example, assuming that the expected TeV flux scales with the spin-down luminosity divided by the distance ($\dot E/{d^2}$), or by the GeV flux observed by Fermi-LAT, have been explored in previous HAWC studies \cite{hawc_collaboration_absence_2025,Albert:2025gwm}. However, in these previous works, no significant differences have been found comparing the results to that obtained with an equal-weighting scheme. Given the current uncertainties on the physical mechanisms of high-energy emission from the spider MSPs, we choose this equal-weighting scheme for the stacking analysis to maintain a model-independent approach.

\section{Results}\label{sec:results}
\subsection{Minimal source fits}

Table~\ref{tab:paras_psr} presents the results of the individual fits. The TS are all below $4$, indicating that there is no firm detection, except for three sources: J1952+2630, J1957+2516, and B1957+20 (or PSR J1959+2048, the first \textit{black widow} ever identified \cite{fruchter_millisecond_1988}). For these, this simple fit returns a TS of 23.67, 4.08, and 5.22, respectively. We therefore perform a more detailed analysis through the multi-source fitting as described in Section~\ref{method:multi} and present the results in Section~\ref{sec:multiResults}. We replace the initial simple fit results of the three sources in Table~\ref{tab:paras_psr} with multi-source fitting results.

\begin{figure*} 
\centering
\includegraphics[width = \textwidth]{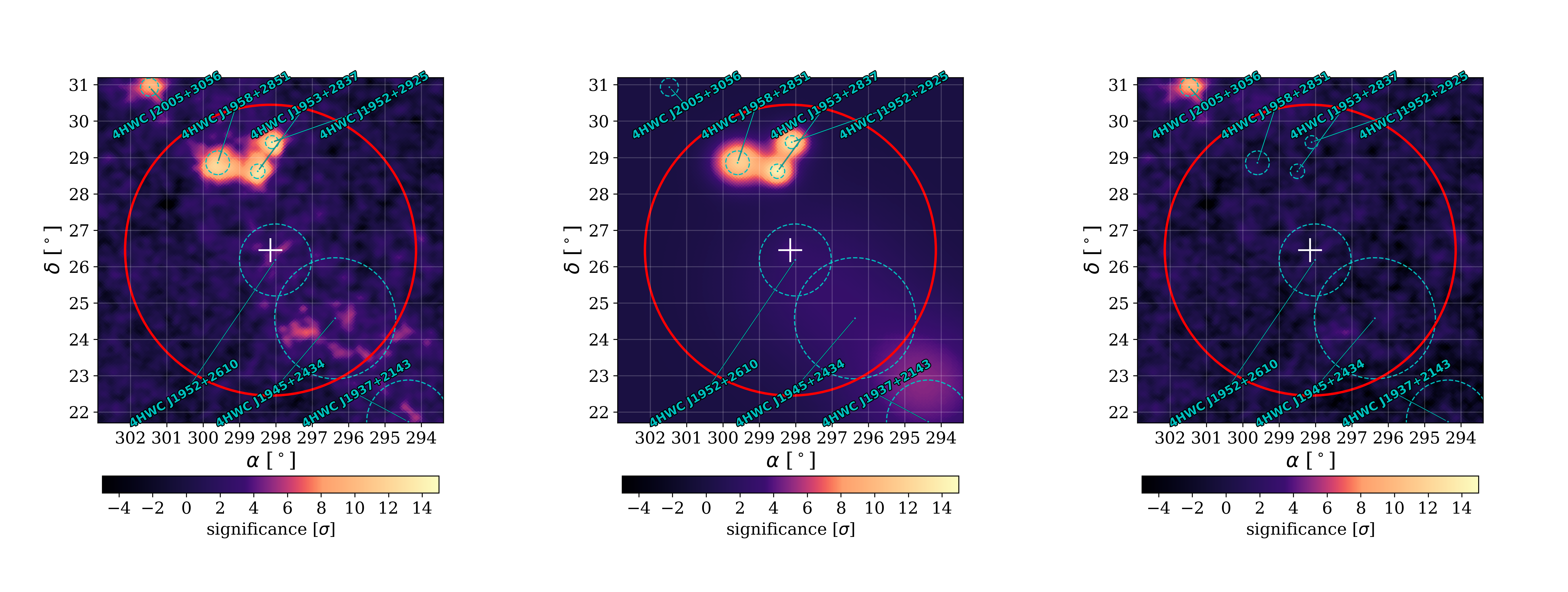}
\caption{Significance maps of the data (left), model (middle) and residual emission (right) in the region around PSR J1952+2630 Fig 3. The fit uses a circular ROI of radius $4.0 ^\circ$ centered at the pulsar, indicated by a red circle. A faint extended source, 4HWC J1945+2434, and three bright sources with small extensions (4HWC J1958+2851, 4HWC J1953+2837, 4HWC J1952+2925) are included in the multi-source fit. The white cross indicates the location of PSR J1952+2630. The dashed lines represent the extensions of the 4HWC sources.}
\label{fig:J1952+2630}
\end{figure*}

\begin{figure*} 
\centering
\includegraphics[width = \textwidth]{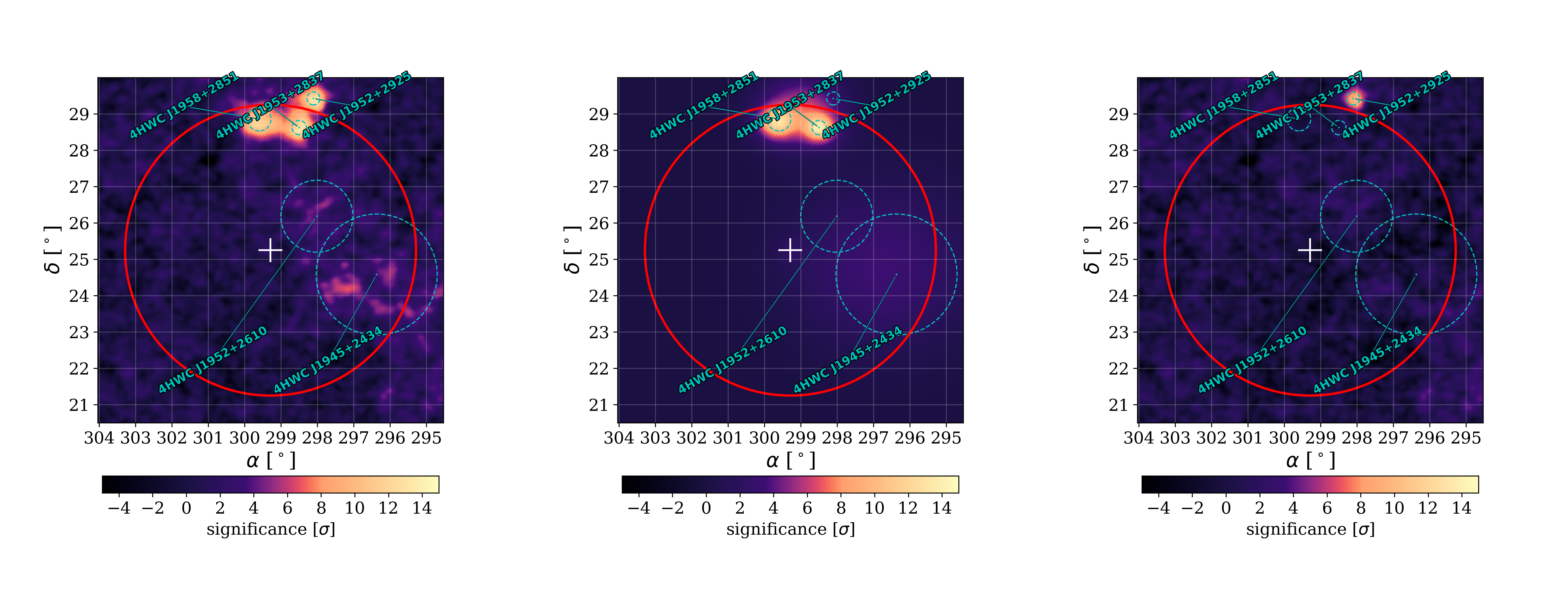}
\caption{Same as Figure~\ref{fig:J1952+2630} but for the region around PSR J1957+2516. This multi-source fit uses a circular ROI of radius $4.0 ^\circ$ (indicated by a red circle) centered at PSR J1957+2516 and includes 4HWC J1945+2434, 4HWC J1958+2851, 4HWC J1953+2837, and 4HWC J1952+2925. The white cross at the center represents the location of PSR J1957+2516.}
\label{fig:J1957+2516}
\end{figure*}

\begin{figure*} 
\centering
\includegraphics[width = \textwidth]{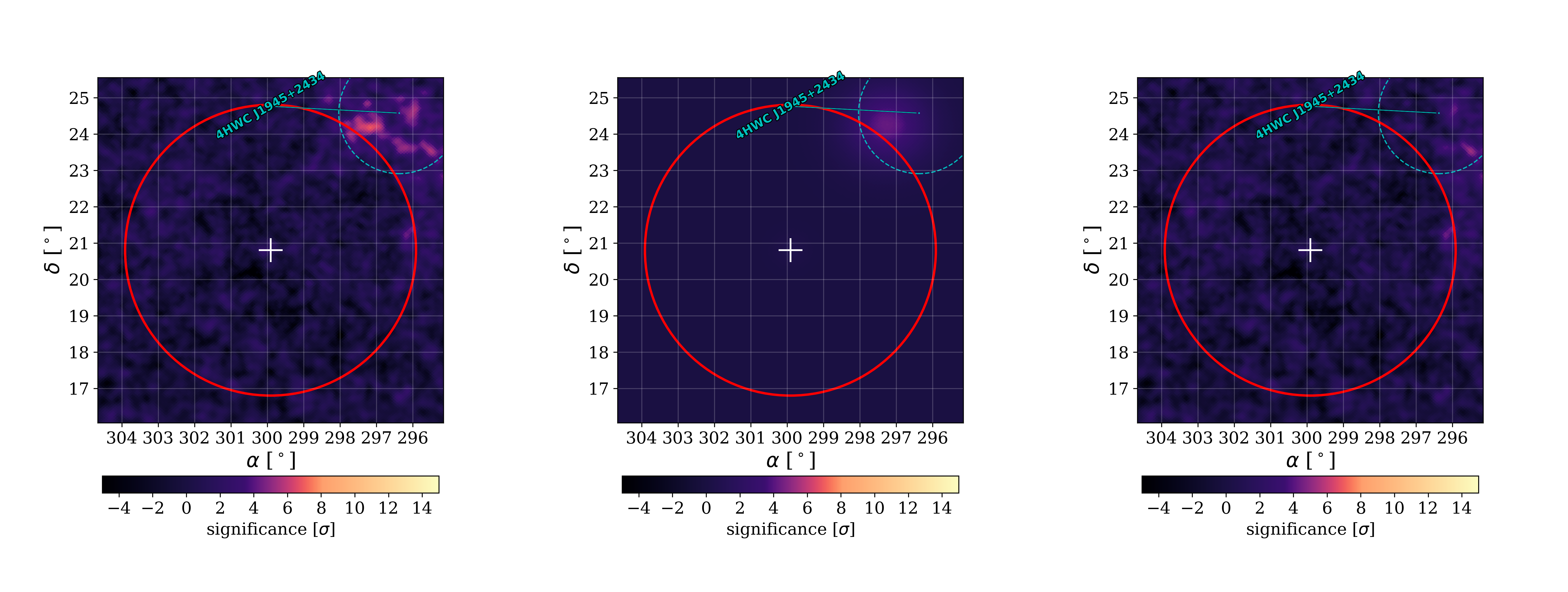}
\caption{Same as Figure~\ref{fig:J1952+2630} but for the region around PSR B1957+20.  This fit  uses a circular ROI of radius $4.0 ^\circ$ (indicated by a red circle) centered at PSR B1957+20 and includes 4HWC J1945+2434. The white cross at the center represents the location of PSR B1957+20.}
\label{fig:B1957+20}
\end{figure*}

\subsection{Multi-source fits}\label{sec:multiResults}

In our minimal source analysis, we fit one source to the ROI and obtain the TS of such a single-source model. This simple study returned three sources with TS higher than 4, suggesting potential excess emission in the region. 
For each of them, we perform multi-source fitting and summarize the key components of the models used to fit the data of the three regions.

Figure~\ref{fig:J1952+2630} shows the significance maps of the data, model and residual emission in the region around PSR J1952+2630. 4HWC J1952+2610, located close to PSR J1952+2630 and likely the same source or associated, is excluded from the fit. Other sources detected in the region include four with smaller extension and high significance: 4HWC J2005+3036, 4HWC J1958+2851, 4HWC J1953+2837, and 4HWC J1952+2925, and two with larger extensions: 4HWC J1945+2434 and 4HWC J1937+2143. The two extended sources are likely the main cause of the high TS obtained without a multi-source fit. We excluded 4HWC J2005+3036 in the fit since it is very close to the ROI boundary and causes parameter overflow during the multi-source fit. When including five extra sources in the ROI, PSR J1952+2630 has a best-fit $TS=0$, which is consistent with background fluctuations. The residual map shows that the significance at the location of PSR J1952+2630 can be well explained by the nearby extended sources rather than a point-like source at the position of the pulsar.

PSR J1957+2516 is located in a similar region as PSR J1952+2630, so the ROIs of the two sources largely overlap. The sources in the fit are mostly the same, except that 4HWC J1952+2925 is now on the ROI boundary. Figure \ref{fig:J1957+2516} shows the model map and residual map of the fit. Similar to PSR J1952+2630, PSR J1957+2516 has a best-fit TS of $2.19\times 10^{-4}$. The residual map also shows that the significance at its location can be explained by the nearby extended sources. Neither location shows significant emission when multi-source fitting is applied.

PSR B1957+20 is further away from the crowded region containing the previous two sources shown in Figure~\ref{fig:J1952+2630}, and only 4HWC J1945+2434 is within its ROI. Figure \ref{fig:B1957+20} shows the model map and residual map of the fit. The best-fit $TS=0$ is also consistent with the background. The moderate TS obtained in the simple fit is likely a result of the extended source 4HWC J1945+2434.  

The TS of all three sources are consistent with zero as reported in Table~\ref{tab:paras_psr}. We therefore conclude that the relatively high individual TS values obtained during the minimal source analysis are likely due to contamination from nearby sources.

\subsection{Stacking analysis}

Table~\ref{tab:stacked_results} presents the results from the stacking of the entire source population and their subsets, including redbacks and black widows. 
The stacking analysis also found a TS that is consistent with background fluctuations.

\begin{figure*}[t]
\centering
\includegraphics[width = \textwidth]{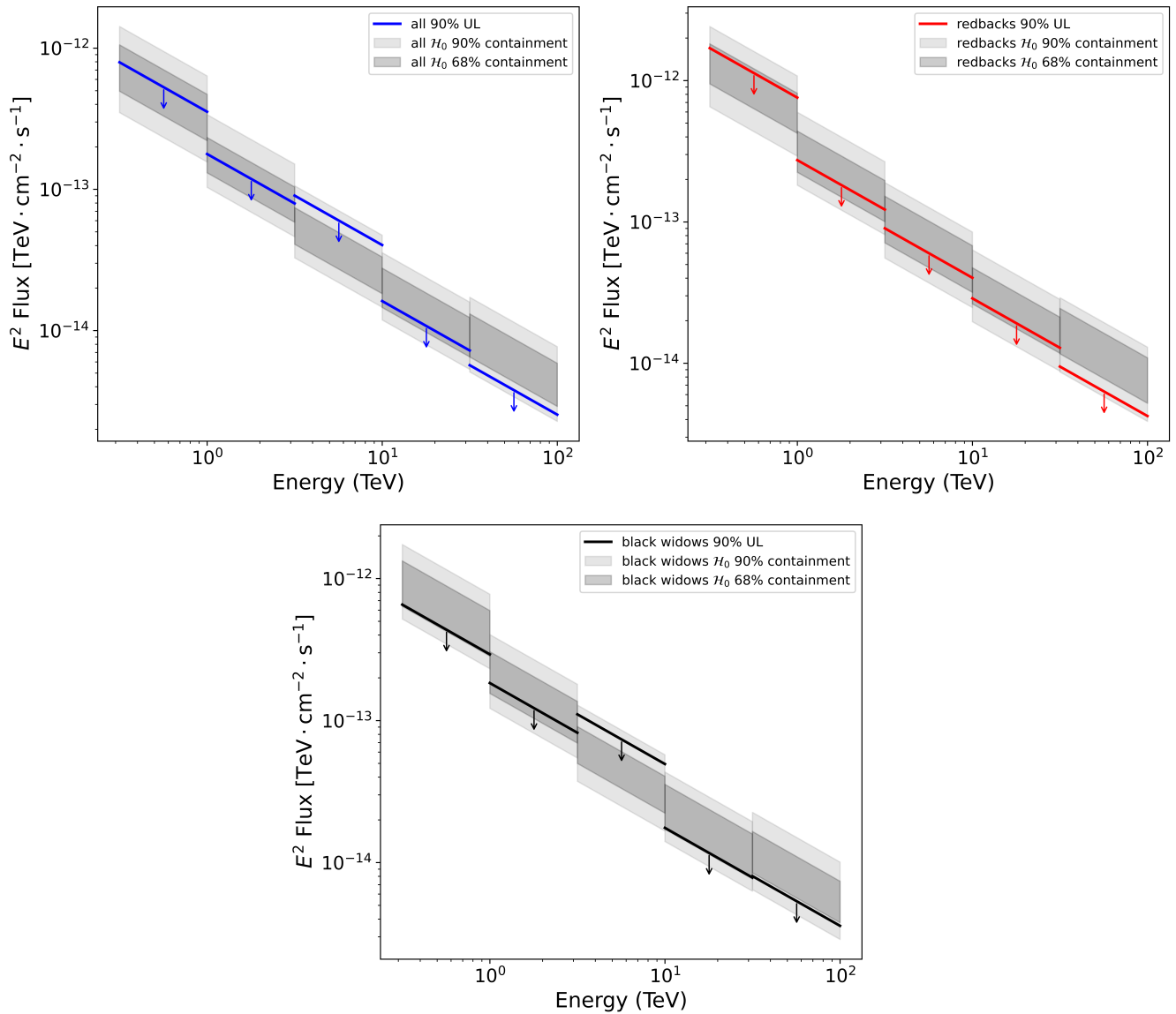}
\caption{\label{fig:gam} Upper limits as a function of energy on the  average flux of MSPs in our sample. The dark and light gray bands correspond to the 68\% and 90\% containment, respectively, for expected upper limits obtained by performing the same analysis but with simulated background data ($\mathcal{H}_0$). We report the results for the full set, the redbacks only, and the black widows only in the upper-left, upper-right, and lower panels, respectively.}
\label{fig:flux}
\end{figure*}

We present the upper limits on the average flux of the sources in Figure~\ref{fig:flux}. The average flux is computed by dividing the total stacked flux by the number of sources in a sample. The grey shaded regions, also referred to as the Brazil bands, present the sensitivity of the study for comparison.  
These are computed by replacing the actual data with Poisson-fluctuated background realizations and performing the analysis using the same source locations and models. This procedure is repeated 100 times and the 68\% and 90\% percentiles of the resulting distributions are used to construct the grey uncertainty bands. Figure \ref{fig:flux} indicates that our upper limits are consistent with the detector’s sensitivity. No significant difference in TeV emission between the redback and black widow groups can be identified given the current sensitivity.

\begin{table}
\centering
\begin{tabular}{ccc}
\hline
\multicolumn{1}{c}{Energy Bin} & \multicolumn{1}{c}{Total TS} & \multicolumn{1}{c}{90\% UL} \\
\multicolumn{1}{c}{} & \multicolumn{1}{c}{} & \multicolumn{1}{c}{[$1 / \rm (cm^2\cdot TeV\cdot s)$]} \\\hline
\multicolumn{3}{c}{\textbf{all}}\\\hline
full & 0.29 & $3.47 \times 10^{-15}$ \\
1 & 0.28 & $6.87 \times 10^{-11}$ \\
2 & 0.22 & $1.54 \times 10^{-12}$ \\
3 & 3.63 & $7.82 \times 10^{-14}$ \\
4 & $1.74 \times 10^{-5}$ & $1.37 \times 10^{-15}$ \\
5 & 0.00 & $4.92 \times 10^{-17}$ \\\hline
\multicolumn{3}{c}{\textbf{redbacks}}\\\hline
full & 0.00 & $1.48 \times 10^{-15}$ \\
1 & 0.98 & $4.66 \times 10^{-11}$ \\
2 & 0.00 & $7.52 \times 10^{-13}$ \\
3 & 0.01 & $2.47 \times 10^{-14}$ \\
4 & $3.08 \times 10^{-4}$ & $7.90 \times 10^{-16}$ \\
5 & $7.74 \times 10^{-5}$ & $2.60 \times 10^{-17}$ \\\hline
\multicolumn{3}{c}{\textbf{black widows}}\\\hline
full & $2.27 \times 10^{-4}$ & $2.45 \times 10^{-15}$ \\
1 & $1.65 \times 10^{-4}$ & $3.81 \times 10^{-11}$ \\
2 & $9.55 \times 10^{-5}$ & $1.04 \times 10^{-12}$ \\
3 & 3.73 & $6.39 \times 10^{-14}$ \\
4 & $4.66 \times 10^{-6}$ & $1.04 \times 10^{-15}$ \\
5 & $3.50 \times 10^{-6}$ & $4.43 \times 10^{-17}$ \\ \hline
\end{tabular}
\caption{Test statistics and 90\% upper limits obtained from the stacking analysis using all pulsars, redbacks, and black widows in the five energy bins. Upper limit values denote the differential flux at the pivot energy of each bin (see Table~\ref{tab:paras_e}). The normalization for the full energy range range is obtained from a fit across the full energy range 0.32–100 TeV.}
\label{tab:stacked_results}
\end{table}

\section{Systematic Uncertainties}\label{sec:syst}

\begin{table*}
\centering
\begin{tabular}{c|ccc|ccc|ccc}
\hline
& & \textbf{all} &  & & \textbf{redbacks} & &  & \textbf{black widows} &  \\
Energy Bin & Index & Response & Total & Index & Response & Total & Index & Response & Total \\
\hline
1 & $+4.99\%$ & $+19.72\%$ & $+20.34\%$ & $+5.36\%$ & $+18.67\%$ & $+19.42\%$ & $+4.99\%$ & $+19.72\%$ & $+20.34\%$ \\
 & $-7.30\%$ & $-6.53\%$ & $-9.79\%$ & $-5.18\%$ & $-6.65\%$ & $-8.43\%$ & $-7.30\%$ & $-6.53\%$ & $-9.79\%$ \\\hline
2 & $+4.47\%$ & $+33.52\%$ & $+33.82\%$ & $+7.85\%$ & $+27.42\%$ & $+28.52\%$ & $+4.47\%$ & $+33.52\%$ & $+33.82\%$ \\
 & $-2.79\%$ & $-6.70\%$ & $-7.26\%$ & $-3.53\%$ & $-8.99\%$ & $-9.65\%$ & $-2.79\%$ & $-6.70\%$ & $-7.26\%$ \\\hline
3 & $+2.06\%$ & $+6.53\%$ & $+6.85\%$ & $+0.34\%$ & $+8.11\%$ & $+8.12\%$ & $+2.06\%$ & $+6.53\%$ & $+6.85\%$ \\
 & $-5.73\%$ & $-25.89\%$ & $-26.51\%$ & $-5.41\%$ & $-22.64\%$ & $-23.27\%$ & $-5.73\%$ & $-25.89\%$ & $-26.51\%$ \\\hline
4 & $+1.73\%$ & $+4.62\%$ & $+4.94\%$ & $+14.41\%$ & $+4.45\%$ & $+15.08\%$ & $+1.73\%$ & $+4.62\%$ & $+4.94\%$ \\
 & $-13.29\%$ & $-31.21\%$ & $-33.93\%$ & $-0.00\%$ & $-24.58\%$ & $-24.58\%$ & $-13.29\%$ & $-31.21\%$ & $-33.93\%$ \\\hline
5 & $+3.69\%$ & $+3.37\%$ & $+4.99\%$ & $+4.63\%$ & $+1.85\%$ & $+4.99\%$ & $+3.69\%$ & $+3.37\%$ & $+4.99\%$ \\
 & $-3.37\%$ & $-27.08\%$ & $-27.29\%$ & $-1.85\%$ & $-30.25\%$ & $-30.30\%$ & $-3.37\%$ & $-27.08\%$ & $-27.29\%$ \\\hline
\end{tabular}
\caption{Systematics for each stacked fit. The index columns represent systematic uncertainties in the 90\% upper limits associated with variations in the spectral index, while the   response columns account for systematics arising from detector response. }
\label{tab:syst}
\end{table*}

Spectral index and detector performance are two key factors of systematic uncertainty. Variations in the assumed spectral index can affect the flux normalization and significance, especially across broad energy ranges. Additionally, we estimate systematic uncertainties by simulating a range of reasonable detector models to account for components with insufficient understanding in HAWC's event reconstruction stage. Different detector models  with changes in calibration or event reconstruction can introduce discrepancies in the measured flux and spatial distribution of sources. We use the alternate detector response modeling files created using this approach on the same models and fitting procedures to calculate systematics. For detailed descriptions of the detector response files and modeling variations, see \cite{albert_performance_2024}. We thus test the robustness of our results against changes in the spectral index value used to extract the significance for the MSPs' emission, and against the detector response. 

The results of the systematics tests are shown in Table~\ref{tab:syst}. The first 2 columns of systematics using each sample set are obtained by varying the spectral index (among $-2.2, -2.4, -2.6, -2.8, -3.0$) and detector response. Systematics are calculated as the percent difference between the baseline case and the maximum (upper) or minimum (lower) fitted results obtained from the respective parameter variations. The total systematics are determined by taking the quadrature sum of all systematic groups. In bins 1 and 2, the spectral index contributes more to the systematics, while in the remaining three bins, the detector response contributes more. Here we show that the total systematic uncertainty is bounded by $\pm 40\%$. This is twice the overall flux systematic error in the observation of the Crab Nebula ($\pm 20\%$) \cite{albert_performance_2024}, which is within an acceptable range for full sky searches.

\section{Discussion and conclusion}\label{sec:conclusions}

In this work, we analyzed 2,565 days of HAWC data to look for VHE emission from a specific class of MSPs, namely MSP binaries with low-mass companions and compact orbits: \textit{spider} MSPs.
Indeed, several theoretical models suggest that these systems may be efficient particle accelerators through the intra-binary shocks between the pulsar and the companion star~\cite{Wadiasingh:2021wcj,Sim:2024kxi,Richard-Romei:2024nje}.

Accounting for available radio and multi-wavelength information and HAWC sky coverage, we selected a sample of 43 MSPs: 15 redbacks and 28 black widows.
We ran individual source fits at the MSP positions, resulting in null detections for all sources. To assess the evidence for three MSPs with the highest TS values in a minimal source fit, we performed a multi-source fit that also accounts for nearby TeV sources. None of the 43 MSPs in our sample shows emission in the energy range of the analysis, 0.3--100 TeV. For each source, we derived the 90\% upper limits of their gamma-ray flux. 

We looked for the cumulative emission from three different source samples (the full sample and the redbacks and black widows separately) by evaluating the TS of their joint likelihood. As no evidence for stacked emission was found, we set 90\% upper limits on the cumulative emission as a function of energy.

It is worth illustrating how HAWC's upper limits connect to lower-energy \textit{Fermi}-LAT data, and other TeV gamma-ray telescope sensitivities and observations, when available. While data from the outrigger array are not included in this analysis, its future integration is expected to substantially enhance HAWC’s sensitivity above multi-TeV energies. These detectors extend the instrumented area by a factor of about 4–5, thereby increasing the number of well-reconstructed events at the highest energies. As a result, future analyses with a longer time span of data and incorporating the outrigger array may achieve better flux limits and increase the possibility of a detection.

We discuss the three objects, for which a TS higher than 4 suggested a potential excess emission during the minimal source analysis. No excess was however found to be significant after investigating them carefully with a multi-source fitting of the excess region. 

The two redbacks PSR J1952+2630, a 20.7~ms pulsar in orbit with a massive white dwarf companion in a close orbit of 0.392 days, and PSR J1957+2516, a 3.96~ms pulsar hosting a companion in a 0.238 day orbit, are not found as gamma-ray pulsars in the updated \textit{Fermi}-LAT pulsar catalog, the 3PC \cite{Fermi-LAT:2023zzt}. 
The source B1957+20 (J1959+2048), the first MSP discovered within a binary system belonging to the class of black widows, is instead detected at gamma-ray energies by the \textit{Fermi}-LAT with an integrated phase-averaged energy flux (obtained from the spectral fitting) in the 0.1–100 GeV energy band of $G_{100}=1.57\times10^{-11}$~erg cm$^{-2}$ s$^{-1}$, and has been investigated at TeV energies by MAGIC \cite{MAGIC:2017iit}, finding no significant emission.

Interestingly, and similarly to what was found for B1957+20 using HAWC data in  \cite{hawc_collaboration_absence_2025}, our stacking upper limits start to approach optimistic model predictions for inverse Compton emission at the source intrabinary shock \cite{vanderMerwe:2020oun}. 
When compared to the sensitivity of other TeV telescopes, our upper limits provide the leading constraints within the full energy range tested, e.g., surpassing the expected sensitivity from 50h of H.E.S.S. observations by almost an order of magnitude at 10~TeV~\cite{hawc_collaboration_absence_2025}.

The upper limit on the average spider emission in the lowest energy bin, 0.32--1.00~TeV, is found to be $6.87 \times 10^{-13}\ \rm TeV^{-1}cm^{-2}s^{-1}$. 
By inspecting the spectral energy distributions available within the 3PC \textit{Fermi}-LAT catalog for the available sources in our sample (an entry in the catalog is found for about 30 sources among the ones listed in Table~\ref{tab:paras_psr}), we find that upper limits are reported in the 3PC highest energy bin (100~GeV--1~TeV) for all the sources, with the notable exception of the source J1023+0038, which was recognized to belong to the class of transitional MSPs, binary pulsars switching between rotation-powered and accretion-powered states and emitting gamma rays \cite{Stappers_2014,Papitto:2020llj}.

The $2\sigma$ upper limits which can be derived from the results listed in the 3PC in the highest energy bin, are found to range, depending on the source, from $0.3$ to $1.2-1.9~\times 10^{-12}$ TeV/cm$^2$/s, and thus compete with our upper limit on the average emission shown in Figure~\ref{fig:flux}. However, at higher energies, the results of the \textit{Fermi}-LAT are very limited by statistics. Starting from energies higher than 1~TeV, our upper limits nicely extend those reported by the \textit{Fermi}-LAT and further test the potential high-energy emission from spider MSPs, testing, e.g., a simple-power law extrapolation of the last \textit{Fermi}-LAT data points at hundreds of GeV.

When compared with available models for individual redbacks and black widows \cite{vanderMerwe:2020oun,Wadiasingh:2021wcj,Sim:2024kxi}  (see also Appendix of \cite{hawc_collaboration_absence_2025} for illustrative comparisons for the sources in the HAWC field of view, B1957+20, and J2339-0533), our HAWC upper limits on the average flux of spider MSPs are generally found to lie above model predictions for the inverse Compton gamma-ray emission by factors ranging from 10\% to more than one order of magnitude. 
However, upper limits on the single-source or average spectral energy distribution at TeV represent important guidance for the development of phenomenological models of these systems and can constrain crucial parameters.
Specifically, spider source parameters (injection spectral index, target radiation field in the companion) and intrabinary shock properties (magnetic field, shock efficiency, geometry, and inclination angles) are so far poorly constrained, and an improvement of 20--50\% in sensitivity with respect to our derived upper limits could potentially lead to a detection in some optimistic configurations. 

Following the recent estimates of a potential hadronic emission in populations of spider MSPs developed in \cite{Vecchiotti:2025qix}, the average predicted flux for redbacks and black widows at TeV energies is still not constrained by our upper limits, and would be even difficult to reach for the upcoming CTAO \cite{Lopez-Oramas:2025vld}. 
Nevertheless, if protons are efficiently accelerated at the intrabinary shock, the companion magnetic field is sufficiently strong (${\sim}10^3$~G), and the spectrum at the shock is hard enough, our upper limits test some extremal realizations. 

Understanding the possible TeV emission of this sub-population of MSPs also has strong implications for the interpretation of a number of phenomena in which MSPs likely play an important role, such as the interpretation of the Galactic Center GeV excess and its high-energy tail \cite{Manconi:2024tgh,Gautam:2021wqn,Macias:2021boz}, the interpretation of gamma-ray emission in globular clusters \cite{Song:2021zrs}, and the origin of the Galactic TeV diffuse emission \cite{Yan:2023uxd,Dekker:2023six}. 

While we wait for the substantial improvement in sensitivity offered by the upcoming generation of gamma-ray telescopes \cite{CTAConsortium:2023tdz,Celli:2024cny}, deeper observations with existing telescopes and careful theoretical studies targeting the population of the most promising objects will guide the effort to unveil high-energy gamma-ray emission in this promising class of binary MSPs.

\section{Acknowledgements}
SM acknowledges the support of the French Agence Nationale de la Recherche (ANR) under the grant ANR-24-CPJ1-0121-01. SM and FC acknowledge  support of the European Union's Horizon Europe research and innovation program for support under the Marie Sklodowska-Curie Action HE MSCA PF–2021,  grant agreement No.10106280, project \textit{VerSi}.
We acknowledge the support from: the US National Science Foundation (NSF); the US Department of Energy Office of High-Energy Physics; the Laboratory Directed Research and Development (LDRD) program of Los Alamos National Laboratory; Consejo Nacional de Ciencia y Tecnolog\'{i}a (CONACyT), M\'{e}xico, grants LNC-2023-117, 271051, 232656, 260378, 179588, 254964, 258865, 243290, 132197, A1-S-46288, A1-S-22784, CF-2023-I-645, CBF2023-2024-1630, c\'{a}tedras 873, 1563, 341, 323, Red HAWC, M\'{e}xico; DGAPA-UNAM grants IG101323, IN111716-3, IN111419, IA102019, IN106521, IN114924, IN110521 , IN102223; VIEP-BUAP; PIFI 2012, 2013, PROFOCIE 2014, 2015; the University of Wisconsin Alumni Research Foundation; the Institute of Geophysics, Planetary Physics, and Signatures at Los Alamos National Laboratory; Polish Science Centre grant, 2024/53/B/ST9/02671; Coordinaci\'{o}n de la Investigaci\'{o}n Cient\'{i}fica de la Universidad Michoacana; Royal Society - Newton Advanced Fellowship 180385; Gobierno de España and European Union-NextGenerationEU, grant CNS2023- 144099; The Program Management Unit for Human Resources \& Institutional Development, Research and Innovation, NXPO (grant number B16F630069); Coordinaci\'{o}n General Acad\'{e}mica e Innovaci\'{o}n (CGAI-UdeG), PRODEP-SEP UDG-CA-499; Institute of Cosmic Ray Research (ICRR), University of Tokyo. H.F. acknowledges support by NASA under award number 80GSFC21M0002. C.R. acknowledges support from National Research Foundation of Korea (RS-2023-00280210). We also acknowledge the significant contributions over many years of Stefan Westerhoff, Gaurang Yodh and Arnulfo Zepeda Dom\'inguez, all deceased members of the HAWC collaboration. Thanks to Scott Delay, Luciano D\'{i}az and Eduardo Murrieta for technical support.

\bibliographystyle{apsrev4-2}
\bibliography{HAWC_MSP}{}

\end{document}